\documentclass[twocolumn,prl,showpacs,preprintnumbers]{revtex4}
\usepackage{graphicx}
\usepackage{bm}
\usepackage{amsmath,amssymb,epsfig,color}

\usepackage{color}%
\definecolor{grey}{rgb}{0.75,0.75,0.75}
\definecolor{orange}{rgb}{1.0,0.5,0.5}
\definecolor{brown}{rgb}{0.5,0.25,0.0}
\definecolor{pink}{rgb}{1.0,0.5,0.5}

\newcommand{\emdash}{---}

\begin{document}

\title{Hund's paradox and the collisional stabilization of chiral molecules}

\author{Johannes Trost}
\author{Klaus Hornberger}
\affiliation{Arnold Sommerfeld Center for Theoretical Physics, 
Ludwig-Maximilians-Universit{\"a}t M{\"u}nchen, Theresienstra{\ss}e 37, 
80333 Munich, Germany
}

\preprint{Phys. Rev. Lett. \textbf{103}, 023202 (2009)}

\begin{abstract}
We identify the dominant collisional decoherence mechanism which serves to
stabilize and super-select the configuration states of chiral molecules. A high-energy
description of this effect is compared to the results of the exact molecular scattering problem, obtained by solving the coupled-channel equations. It allows to predict the experimental conditions for observing the collisional suppression of the tunneling dynamics between the left- and the right-handed configuration of D$_2$S$_2$ molecules.
\end{abstract}


%
%

\maketitle

{{\em Introduction.{\emdash}\/}}An old problem in molecular quantum mechanics,
first discussed by F. Hund {\cite{Hund1927a}}, is how to explain from first
principles why molecules often appear as {{\em enantiomers\/}}, i.e., either
in a left-handed configuration or as the right-handed mirror image. Given the
parity-invariant molecular Hamiltonian, one might rather expect them in the
ground state, corresponding to the symmetric superposition of these chiral
states. Traditionally, this is explained by the possibly very long tunneling
time from a left-handed configuration state $|L \rangle$ to a right-handed one
$|R \rangle$. However, this does not solve the `paradox' since one still needs
to understand the seeming failure of the superposition principle, prohibiting
superposition states of the form $| \psi_{\xi} \rangle = \left( |L \rangle +
e^{i \pi \xi} |R \rangle \right) / \sqrt{2}$ with $0 \leqslant \xi < 2$
{\cite{Simonius1978a,Harris1981Harris1982}}.

While this super-selection phenomenon has been linked with fundamental parity
violations {\cite{Quack2002a}}, a very natural explanation is offered by the
concept of environmental decoherence {\cite{JoosZurekSchlosshauer}}. What
selects the enantiomer states according to this theory is the fact that the
typical interaction with environmental degrees of freedom, such as the
collision with a gas particle, can better distinguish the alternatives $|L
\rangle$ and $|R \rangle$ than, say, between the molecular eigenstates $|
\psi_0 \rangle$ and $| \psi_1 \rangle$. The enantiomer states are then
prevented both from tunneling between each other and from decaying into a
mixture, if these environmental interactions are sufficiently frequent
compared to the tunneling period. This stabilization can be understood in
analogy to the quantum Zeno effect if one views the environment as
continuously monitoring the molecular state. Superposition states $|
\psi_{\xi} \rangle$, in contrast, would get quickly decohered by their ensuing
quantum correlation with the environment.

This environmental distinction of specific molecular configurations is one of
the paradigms in the field of decoherence, discussed by many of the
path-breaking works {\cite{Simonius1978a,Harris1981Harris1982,ZurekJoos}}. At
the same time, not much is known about the concrete microscopic mechanism at
work with realistic molecules, nor whether the transition from the tunneling
regime to stabilization can be observed experimentally.

In this letter, we use molecular scattering theory to identify the dominant
microscopic mechanism responsible for chiral stabilization due to a background
gas. We show that this effect is determined by a parity-sensitive higher-order
term in the dispersive interaction, which is usually disregarded because it
does not affect the equilibrium properties of gases (due to the orientational
averaging involved). In spite of this, and in spite of the fact that this
process can not be used to separate a racemic mixture into chiral components,
it provides a surprisingly efficient channel for environmental decoherence.
This is demonstrated numerically by solving the full-fledged coupled-channel
problem for a simple chiral molecule, D$_2$S$_2$. Our numerical results
motivate and confirm a high-energy approximation, which allows to assess the
decoherence effect at room temperature and to predict when the
collision-induced transition from tunneling to stabilized chiral states can be
observed experimentally.

{{\em Master equation for collisional stabilization.\/}}{---}The following
microscopic analysis is facilitated by the recent derivation of a master
equation yielding the incoherent dynamics of the internal-rotational molecular
state due to the collisions with a lighter, thermalized background gas
{\cite{Dumcke1985a,Hornberger2007b}}. Crucially, this Markovian description
incorporates the interaction between molecule and gas particle in a
non-perturbative fashion, by means of the multi-channel scattering amplitudes.

We assume that the kinetic energy transferred in the collisions is
sufficiently small compared to both the Born-Oppenheimer barrier separating
the enantiomer states and to the excitation energies of the electronic and
vibrational internal states of the molecule and the gas particle. This is
valid in most cases of interest, and it implies that thermally induced
transitions between $|L \rangle$ and $|R \rangle$ do not occur. It is then
justified to take the interaction operator to be diagonal in the enantiomer
basis, $\hat{V} = V_L \left( \hat{\ensuremath{\boldsymbol{r}}} \right) |L
\rangle \langle L| + V_R \left( \hat{\ensuremath{\boldsymbol{r}}} \right) |R
\rangle \langle R|$, thus restricting the molecular configuration state to a
two-dimensional subspace. Here, $\hat{\ensuremath{\boldsymbol{r}}}$ is the
inter-particle position operator in the body-fixed molecular system; for
simplicity we disregard a possible dependence of $\hat{V}$ on the orientation
of the environmental gas particle.

The master equation {\cite{Hornberger2007b}} simplifies considerably under
these assumptions. It is formulated in terms of the scattering amplitudes for
the channels corresponding to the molecular eigenstates $| \psi_0 \rangle$ and
$| \psi_1 \rangle$. However, since the S-matrix for $\hat{V}$ does not couple
subspaces of different chiral configurations, $\langle L| \hat{S} |R \rangle =
\langle R| \hat{S} |L \rangle = 0$, one can express the proper scattering
amplitudes as linear combinations of the scattering amplitudes $f^{(L)}$ and
$f^{(R)}$ associated to the unitary scattering operators $\langle L| \hat{S}
|L \rangle$ and $\langle R| \hat{S} |R \rangle$. The configuration dynamics
then no longer depends on the scattering cross section. Rather, it is
determined by
\begin{eqnarray}
  \eta_{\alpha \alpha_0} \left( v \right) & = & \int \frac{\mathrm{d}
  \ensuremath{\boldsymbol{n}} \mathrm{d} \ensuremath{\boldsymbol{n}}_0}{8 \pi}
  \left| f_{\alpha, \alpha_0}^{(L)} \left( v\ensuremath{\boldsymbol{n}},
  v\ensuremath{\boldsymbol{n}}_0 \right) - f_{\alpha, \alpha_0}^{(R)} \left(
  v\ensuremath{\boldsymbol{n}}, v\ensuremath{\boldsymbol{n}}_0 \right)
  \right|^2, \nonumber\\&& \label{eq:decocross}
\end{eqnarray}
which may be called a decoherence cross section. Similar to a proper partial
cross section, this characteristic area depends on the relative velocity $v$
and on the initial and final internal state of the molecule, labeled by the
multi-indices $\alpha_0$ and $\alpha$. In the present case, these are the
rotation states of an asymmetric top, $\alpha = \left( j, m_j, \tau \right)$,
specified by the total and azimuthal quantum numbers $j$ and $m_j$, and the
pseudo quantum number $\tau$. Importantly, the decoherence cross section
(\ref{eq:decocross}) depends on the phase difference of $f^{(L)}$ and
$f^{(R)}$. It may thus be quite large even if the corresponding scattering
cross sections are identical, $|f^{(L)} |^2 = |f^{(R)} |^2$. The presence of
the background gas also gives rise to a coherent modification of the tunneling
dynamics, described by the characteristic area $\varepsilon_{\alpha \alpha_0}
\left( v \right) = \int \mathrm{d} \ensuremath{\boldsymbol{n}} \mathrm{d}
\ensuremath{\boldsymbol{n}}_0 \ensuremath{\operatorname{Im}}[f_{\alpha,
\alpha_0}^{(L)} \left( v\ensuremath{\boldsymbol{n}},
v\ensuremath{\boldsymbol{n}}_0 \right) f_{\alpha, \alpha_0}^{(R) \ast} \left(
v\ensuremath{\boldsymbol{n}}, v\ensuremath{\boldsymbol{n}}_0 \right)] / 4 \pi
.$

In the master equation the areas $\eta_{\alpha \alpha_0} \left( v \right)$
and $\varepsilon_{\alpha \alpha_0} \left( v \right)$ are multiplied with the
current density of the gas particles to yield the decoherence rate $\gamma$
and the frequency shift $\omega_x$, respectively. Specifically, the
decoherence rate is given by
\begin{eqnarray}
  \gamma & = & n_{\ensuremath{\operatorname{gas}}} \langle v \, \eta \rangle
  \equiv n_{\ensuremath{\operatorname{gas}}} \sum_{\alpha, \alpha_0} w \left(
  \alpha_0 \right)  \int_0^{\infty} \mathrm{d} v \nu \left( v \right) v
  \eta_{\alpha \alpha_0} \left( v \right), \nonumber\\&&
\end{eqnarray}
where $n_{\ensuremath{\operatorname{gas}}}$ is the number density of the
background gas and $\nu \left( v \right)$ its velocity distribution. Here,
taking the average over the rotational state distribution $w \left( \alpha_0
\right)$ is an additional approximation, which is well allowed if the rotation
frequency is much greater than the tunneling frequency, as is usually the
case. The frequency shift $\omega_x = n_{\ensuremath{\operatorname{gas}}}
\langle v \varepsilon \rangle$ is determined by the same average.

The collisional master equation {\cite{Hornberger2007b}} thus reduces to an
equation in the 2d space spanned by the $| \psi_{\xi} \rangle = \left( |L
\rangle + e^{i \pi \xi} |R \rangle \right) / \sqrt{2}$. Denoting the tunneling
frequency by $\omega_z$ and identifying the energy eigenstates $| \psi_0
\rangle$, $| \psi_1 \rangle$ with the eigenvectors of the Pauli matrix
$\hat{\sigma}_z$, it takes the form
\begin{eqnarray}
  \partial_t \rho & = & \frac{1}{2 i} \left[ \omega_z \hat{\sigma}_z +
  \omega_x \hat{\sigma}_x, \rho \right] + \frac{\gamma}{2} \left(
  \hat{\sigma}_x \rho \hat{\sigma}_x - \rho \right) . 
\end{eqnarray}
This equation has the remarkable property of stabilizing the enantiomer states
$|R \rangle$ and $|L \rangle$, provided $\gamma \gg \omega_z$, i.e., when the
decoherence rate is much greater than the tunneling rate. The chiral
configuration states decay with the suppressed rate $\omega_z^2 / \gamma \ll
\gamma, w_z$ in this case, while the superposition states $| \psi_{\xi}
\rangle$ decay at least with rate $\gamma$
{\cite{Simonius1978a,Harris1981Harris1982}}.

{{\em The parity-sensitive dispersion interaction.{\emdash}\/}}We now turn to
the general van der Waals interaction between two polarizable particles to
identify the relevant part that distinguishes a left-handed from a
right-handed configuration and thus gives rise to different scattering
amplitudes $f^{(L)}$ and $f^{(R)}$. The total London dispersion interaction is
conveniently expressed in terms of the frequency-dependent
multipole-polarizability tensors of the particles, most prominently those
involving virtual electric dipole (ED) and electric quadrupole (EQ)
transitions {\cite{Stone96}}. It is dominated by the standard van der Waals
ED-ED/ED-ED interaction, which depends on the inter-particle distance $r$ as
$r^{- 6}$ and is determined by the ED-ED polarizability tensors $\alpha \left(
i \omega \right)$. This bulk interaction cannot distinguish different chiral
configurations since the $\alpha \left( i \omega \right)$ are parity-invariant.

Among the chirally sensitive parts of the London dispersion interaction
between a chiral and an achiral molecule, the most important contribution is
due to the EQ-ED/ED-ED polarizability combination. It has a $r^{-
7}$dependence and it combines the EQ-ED polarizability tensor $A_{i, jk}
\left( i \omega \right)$ of the chiral molecule {\cite{Stone96}} with the
ED-ED polarizability of the spherical projectile. One can safely neglect the
contributions involving higher-order electric polarizability tensors due to
their short-ranged nature; among the contributions involving magnetic susceptibilities, in particular due to magnetic dipole (MD) and magnetic quadrupole (MQ) transitions, the ED-MD/ED-MD vanishes if one molecule is achiral, while
the ED-MD/ED-MQ and ED-MD/EQ-MD are suppressed by the square of the fine
structure constant {\cite{Chiu80,Jenkins1994a}}. Also the (parity-invariant)
influence of a small permanent dipole-moment of the chiral molecules can be
usually disregarded, since it gets overshadowed by the van der Waals
interaction.

We note that the chirally sensitive EQ-ED/ED-ED interaction contribution is
usually not accounted for in molecular scattering calculations. This is
because its chiral disparity vanishes when averaged over all orientations of
the chiral molecule, implying that the cross sections of left- and
right-handed molecules are equal. However, such rotational averaging of the
coupled-channel equations, which drastically reduces the calculational effort,
is not allowed in the present case, since the decoherence cross section
(\ref{eq:decocross}) crucially depends on the phase differences of the
individual scattering amplitudes $f^{\left( L \right)}$ and $f^{\left( R
\right)}$. This phase information is lost at the level of standard scattering
cross sections, which indeed remain equal for left- and right-handed
molecules.

{{\em Solving the coupled-channel equations.{\emdash}\/}}In order to evaluate
the scattering amplitudes one must calculate the scattering matrix $S$ by
solving the associated coupled-channel equations {\cite{Child84}}. Our system
comprises a chiral asymmetric top molecule colliding with a noble gas atom,
such that each channel is described by a set of rotational (pseudo) quantum
numbers $j$, $m_j$, $\tau$, as well as by the orbital and the total angular
momentum, $\ell$ and $J$.

The resulting system of differential equations involves infinitely many closed
channels and a finite number of asymptotically free, open channels, which
scales with the total energy as $E^{3 / 2}$. When truncating this infinite
system, care is needed to ensure that at least all those closed channels are
retained which have an appreciable coupling to the initial state, even if
their asymptotic energies lie deeply in the energetically forbidden domain.

Since state-of-the-art numerical program packages for scattering calculations
do not accommodate asymmetric molecules with their parity-changing collision
dynamics, we developed a numerical method for solving the full quantum
mechanical scattering problem (avoiding premature rotational averaging). It is
based on the log-derivative algorithm of Johnson {\cite{Johnson73}}, and it
allows to ensure the convergence with respect to the truncation of closed
channels by adjusting threshold values both for the closed channel energies
and for the internal angular momenta to be retained. We checked our program
for the case of symmetric top molecules against the numerical package
{\textsc{Molscat}} {\cite{MOLSCATSHORT}}.

{{\em Model for D$_2$S$_2$.{\emdash}\/}}The transition from tunneling to
stabilization is best observed with a molecule of moderate tunneling frequency
$\omega_z$. Motivated by recent proposals for enantiomer discrimination
{\cite{Thanopulos2003a,Li2007a}}, we focus on D$_2$S$_2$, one of the simplest
chiral molecules, with $\omega_z / 2 \pi = 176 \,
\ensuremath{\operatorname{Hz}}$. Since {{\em ab initio\/}} calculations for
the dynamic tensors $\alpha \left( i \omega \right)$ and $A_{i, jk} \left( i
\omega \right)$ are not yet feasible for dihydrogendisulfide (nor for any
other chiral molecule) we extend the bond increment method {\cite{Ewig02}} for
the calculation of static molecular susceptibilities to the dynamic case. As
described in {\cite{Ewig02}}, static molecular susceptibility tensors can be
calculated as a sum over bond increments associated to the constituent atoms
$a$, which account for the relative positions and orientation of the atoms and
their mutual bonds. The bond increments are obtained from {{\em ab initio\/}}
values of static molecular susceptibilities of training compounds, which
include the D$_2$S$_2$ molecule {\cite{Ewig02}}. In a simple extension of this
approach, we replace the static values of the constituent atomic
susceptibilities by a Drude model, $\alpha^{\left( a \right)} \left( i \omega
\right) = \alpha^{\left( a \right)} \left( 0 \right) f_a \left( \omega
\right)$ and $A_{i, jk}^{\left( a \right)} \left( i \omega \right) = A_{i,
jk}^{\left( a \right)} \left( 0 \right) f_a \left( \omega \right)$, with $f_a
\left( \omega \right) = \omega^2_a / \left( \omega^2_a + \omega^2 \right)$,
where $\omega_a$ is the first excitation energy, $\omega_{\text{D}} =
0.375$a.u and $\omega_{\text{S}} = 0.252$a.u. {\cite{cccbdb}}. The
configuration of D$_2$S$_2$ is characterized by the bond lengths
$r_{\ensuremath{\operatorname{SS}}} =$2.05{\AA},
$r_{\ensuremath{\operatorname{SD}}}$=1.34{\AA}, the inter-bond angle $\angle
(\ensuremath{\operatorname{DSS}}) = 100.4^{\circ}$ and the dihedral angle
$\angle (\ensuremath{\operatorname{DSSD}}) = 90.3^{\circ}$, and constants of
inertia given in {\cite{Winnewisser91}}. We take the background gas to be
ground state helium, whose (spherically symmetric) dipole-dipole
polarizability is accurately described by a sum of four Lorentzians
{\cite{chan65}}.\begin{figure}[tb]
  \resizebox{\columnwidth}{!}{\epsfig{file=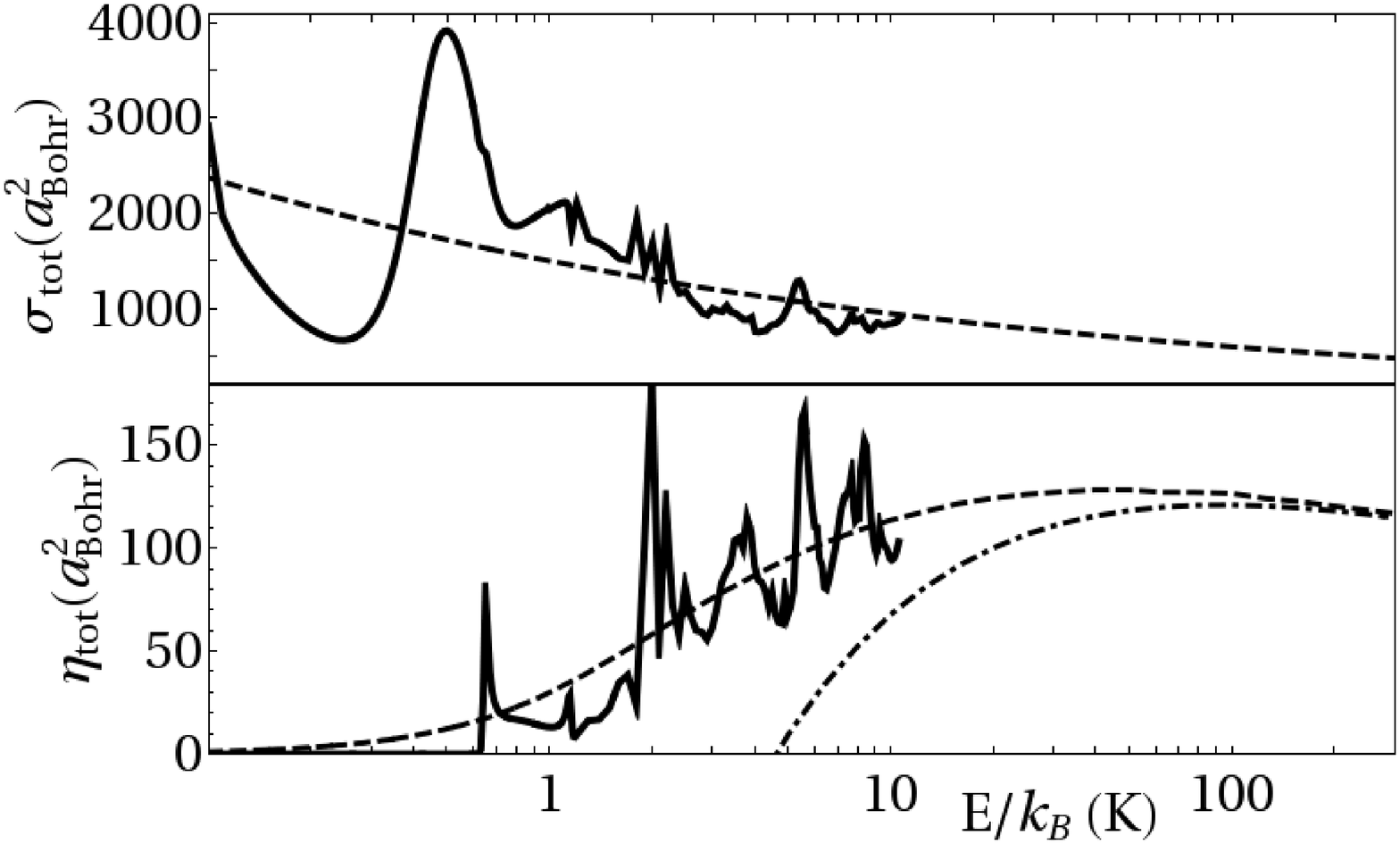}}
  \caption{\label{fig:numerics}Scattering cross section
  $\sigma_{\ensuremath{\operatorname{tot}}}$ (upper panel, solid line) and
  decoherence cross section $\eta_{\ensuremath{\operatorname{tot}}}$ (lower
  panel, solid line) for the collision of a He atom off a ground state
  D$_2$S$_2$ molecule, as a function of the kinetic energy $E / k_B \leqslant
  10$K on a logarithmic scale. The dashed line corresponds to
  (\ref{eq:etaint}), the dash-dotted lines gives the high-energy behavior
  (\ref{eq:asympratio}).}
\end{figure}

{{\em Numerical results.{\emdash}\/}}It takes substantial numerical effort to
observe the onset of collisional stabilization, since this regime is
characterized by a large number of partial waves, while the standard
semiclassical approaches for cross sections cannot be applied. Moreover, the
number of relevant scattering channels proliferates; at the kinetic energy $E
/ k_{\text{B}} = 300 \, \text{K}$ of the relative motion about $1.2 \times
10^4$ coupled differential equations would have to be solved simultaneously,
with as many different initial conditions, for extracting the S-matrix in the
subspace of a single total angular momentum $J$.

The upper panel of Fig. \ref{fig:numerics} presents the exact total
scattering cross section $\sigma_{\ensuremath{\operatorname{tot}}}$ (for
kinetic energies up to 10K), starting from the rotational ground state of
D$_2$S$_2$, while the lower panel shows the decoherence cross section
$\eta_{\ensuremath{\operatorname{tot}}} = \Sigma_{\alpha} \eta_{\alpha
\alpha_0}$. One observes that a non-negligible effect of collisional
stabilization can be expected already for energies well below the threshold
for the first channel that directly couples different parity subspaces
($E_{\ensuremath{\operatorname{thres}}} / k_B = 17.5 \,$K). The decoherence
cross section $\eta_{\ensuremath{\operatorname{tot}}}$ tends to saturate at
about 100$a_0^2$, corresponding to $\eta_{\ensuremath{\operatorname{tot}}} /
\sigma_{\ensuremath{\operatorname{tot}}} \simeq 10 \, \%$ at 10K{---}a
remarkably large value, given that the chirality distinguishing interaction
contributes only weakly to $\sigma_{\ensuremath{\operatorname{tot}}}$.

{{\em High-energy approximation.{\emdash}\/}}In order to understand this
saturation and to assess the stabilizing effect of collisions at larger
temperatures we now consider a high-energy approximation for
$\eta_{\ensuremath{\operatorname{tot}}}$ and
$\sigma_{\ensuremath{\operatorname{tot}}}$. Since the Born approximation
renders the scattering amplitude a linear functional of the potential, it
follows immediately from (\ref{eq:decocross}) that the asymptotic high-energy
behavior of $\eta_{\ensuremath{\operatorname{tot}}}$ is the cross section
corresponding to $\Delta V \left( \ensuremath{\boldsymbol{r}} \right) = V_L
\left( \ensuremath{\boldsymbol{r}} \right) - V_R \left(
\ensuremath{\boldsymbol{r}} \right)$. We start from the general dependence of
the cross section on the initial state elastic S-matrix element, \
$\sigma_{\ensuremath{\operatorname{tot}}} = 2 \pi \sum_J \left( 2 J + 1
\right) (1 -\ensuremath{\operatorname{Re}}(\langle \Psi_0 | \hat{S}^{\left( J
\right)} | \Psi_0 \rangle))  / k^2$ {\cite{Child84}}, apply the exponential
Born approximation, and take the high-$J$ asymptotics. For a homogeneous,
spherical potential $V \left( r \right) = C_n r^{- n}$, this yields the
standard result $\tilde{\sigma}^{\left( n
\right)}_{\ensuremath{\operatorname{tot}}} = p_n \left( C^2_n / E \right)^{1 /
\left( n - 1 \right)}$, with $p_n$ a numerical factor depending weakly on $n >
2$ {\cite{Child84}}.

Disregarding the molecular core, we can thus approximate
$\sigma_{\ensuremath{\operatorname{tot}}}$ by the dominant, spherically
symmetric $n = 6$ contribution to the van der Waals ED-ED/ED-ED interaction,
i.e., $\sigma_{\ensuremath{\operatorname{tot}}} \simeq
\tilde{\sigma}_{\ensuremath{\operatorname{tot}}}^{\left( 6 \right)}$ with $C_6
= 11.7 \text{a.u.}$ Also $\Delta V \left( \ensuremath{\boldsymbol{r}} \right)$
is homogeneous in $r$, with $n = 7$, but there is no direct Born contribution
to the decoherence cross section $\eta_{\ensuremath{\operatorname{tot}}}$
since $\langle \Psi_0 | \Delta V \left( \hat{\ensuremath{\boldsymbol{r}}}
\right) | \Psi_0 \rangle$ vanishes. Therefore, we include all off-diagonal
elements that couple the ground state in the exponential Born expression. This
yields $\ensuremath{\operatorname{Re}}(\langle \Psi_0 | \hat{S}^{\left( J
\right)} | \Psi_0 \rangle) \simeq \cos ( \sqrt{\sum_{f \neq 0} \left| \langle
\Psi_0 | \Delta V| \Psi_f \rangle \right|^2 })$, with $\Psi_f$ the free
channel wave functions in the $J$ subspace.

The lowest channel $\Psi_1 $that couples to $\Psi_0$ via $\Delta V$ opens at
$17.5 K$, with $j = 3$. Expressing the corresponding coupling as $\left|
\langle \Psi_0 | \Delta V| \Psi_1 \rangle \right| = \hbar^2 \beta^5 / \left( 2
m_{\ast} r^7 \right)$, with $\beta$ a convenient parameterization and
$m_{\ast}$ the reduced mass, and replacing the sum by an integral this yields
\begin{eqnarray}
  \eta_{\ensuremath{\operatorname{tot}}} & \cong & \frac{4 \pi}{k^2}
  \int^{\infty}_{1 / 2} \mathrm{d} J \left( 2 J + 1 \right) \sin^2 \left\{
  \frac{5 \pi}{128}  \left( k \beta \right)^5 \frac{\Gamma \left( J -
  \frac{1}{2} \right)}{\Gamma \left( J + \frac{11}{2} \right)}  \right\} . 
  \nonumber\\&&\label{eq:etaint}
\end{eqnarray}
with $E = \left( \hbar k \right)^2 / 2 m_{\ast}$. This approximation is
displayed as the dashed line in the lower panel of Fig.~\ref{fig:numerics}.
The comparison with the exact results indicates that (\ref{eq:asympratio})
starts to apply already at numerically accessible energies. An asymptotic
expansion yields
\begin{equation}
  \eta_{\ensuremath{\operatorname{tot}}} \simeq c_1 \frac{\beta^{5 / 3}}{k^{1
  / 3}} - c_2 \frac{\beta^{5 / 6}}{k^{7 / 6}} \label{eq:asympratio}
\end{equation}
with $c_1 = 3.66$ and $c_2 = 14.4$ (see dash-dotted line in
Fig.~\ref{fig:numerics}).

{{\em Discussion.{\emdash}\/}}Our numerical and analytical analysis suggests
that the ratio $\eta_{\ensuremath{\operatorname{tot}}} /
\sigma_{\ensuremath{\operatorname{tot}}}$ varies only weakly at larger
energies. A typical value at 300\,K of $\eta_{\ensuremath{\operatorname{tot}}}
/ \sigma_{\ensuremath{\operatorname{tot}}} \simeq 25\%$ means that
environmental stabilization sets in once the rate of collisions exceeds four
times the tunneling frequency. Noting that D$_2$S$_2$ tunnels with $\omega_z /
2 \pi = 176 \, \ensuremath{\operatorname{Hz}}$, we find from
(\ref{eq:asympratio}) that the critical pressure and temperature of a helium
atmosphere must satisfy $\left( p /\ensuremath{\operatorname{mbar}} \right)
\left( T / \text{K} \right)^{- 2 / 3} \geqslant 3.0 \times 10^{- 7}$, i.e., $p
\geqslant 1.6 \times 10^{- 5} \ensuremath{\operatorname{mbar}}$ for $T = 300
\, \text{K}$.

This prediction could be tested in an experiment that applies laser based
coherent control techniques in a Stern-Gerlach-type setup for separating a
molecular beam into left- and right-handed daughter beams
{\cite{Thanopulos2003a,Li2007a}}. Passing one of them through a gas cell, one
may analyze the enantiomeric purity with another Stern-Gerlach stage. Since
thermal racemization can be controlled by cooling (given the $2300 \, \text{K}
k_B$ barrier height for chiral flipping in D$_2$S$_2$), it is thus possible to
directly observe the expected environmental stabilization of the chiral
configuration as a function of the gas pressure.

This work was supported by the DFG Emmy Noether program.

\end{document}